# Head-mounted Displays, Smartphones, or Smartwatches? – Augmenting Conversations with Digital Representation of Self

ILYENA HIRSKYJ-DOUGLAS, Aalto University, Finland
MIKKO KYTÖ, Aalto University and University of Helsinki, Finland
DAVID MCGOOKIN, Aalto University, Finland

Technologies that augment face-to-face interactions with a digital sense of self have been used to support conversations. That work has employed one homogenous technology, either 'off-the-shelf' or with a bespoke prototype, across all participants. Beyond speculative instances, it is unclear what technology individuals themselves would choose, if any, to augment their social interactions; what influence it may exert; or how use of heterogeneous devices may affect the value of this augmentation. This is important, as the devices that we use directly affect our behaviour, influencing affordances and how we engage in social interactions. Through a study of 28 participants, we compared head-mounted display, smartphones, and smartwatches to support digital augmentation of self during face-to-face interactions within a group. We identified a preference among participants for head-mounted displays to support privacy, while smartwatches and smartphones better supported conversational events (such as grounding and repair), along with group use through screen-sharing. Accordingly, we present software and hardware design recommendations and user interface guidelines for integrating a digital form of self into face-to-face conversations.

## 1 INTRODUCTION

We use various social-interaction-supporting technologies to augment our communications with each other. This is typically through online media applied for self-representation such as Facebook, LinkedIn, and Instagram and to support online conversations as with Facebook

This is work is supported by the Academy of Finland, under grant 311090.
Author's addresses: I. Hirskyj-Douglas, Aalto University, P.O. Box 11000, FI-00076, Finland; M. Kytö, University of Helsinki, Gustaf Hällströmin katu 2b, 00560, Finland; D. McGookin, Aalto University, P.O. Box 11000, FI-00076, Finland.

Messenger and WhatsApp [12]. Recently, attention has been directed to how people use technology in a collocated application through entwining of the digital and face-to-face interactions [15]. These collocated interactions consist primarily of a digital representation aimed at providing a 'ticket-to-talk' [64] to support social interactions [6, 19–22, 24]. For instance, during a large networking event such as a party or conference, technology could help people pinpoint those to whom they wish to speak first, on the basis of interests, organisation affiliation, and/or experience. This conversation-opener allows for not only a quick connection but also getting a quick overview of the others present. With loneliness being frequently identified as a 21st-century epidemic and as a problem for a substantial number of people, it is hoped that such technologies will help encourage users to engage in meaningful conversations [1, 69].

These virtual mediated social interactions in which people can connect have evolved from teletext adverts into chatrooms, bulletin board systems, online discussion groups, and instant messaging [1]. Recently, an increasing movement has emerged in which technology is used to augment face-to-face connection with others in what Lampe et al. [33] have called social searching. With face-to-face conversations shown to increase our happiness, even with strangers [11], there is ongoing discussion of technology's impact on face-to-face interactions [33] and the quality of these interactions [56]. Paul [1] has highlighted the benefit of viewing people's online profiles with regard to both more intimate conversations and deeper disclosure fostering a sense of trust and intimacy that leads towards friendship. Still, there are unresolved questions about how we can structure and weave in these online representations in connection with face-to-face conversations and the perceived outcomes.

Current technologies to augment and support social interactions have been used within various contexts, from conferences [5, 41] to small-group interaction [32], on a spectrum of prototyping as a public interface [20] to initial targeting at potential face-to-face interactions [38]. These types of augmentations, however, have yet to be deployed in an informal group setting with many users. There are constraints on usage in various contexts, and people might choose different devices in, for example, different places. Thus, an opening exists within the field of augmenting social interaction for examining how current technologies can be implemented and the effects of such use for face-to-face interactions. Doing this is imperative since, in everyday contexts, often conversations are held within a sizeable set of people or at a gathering with a mix of strangers and friends. Yet with research not investigating this normal occurrence at present, little is known of how to map and scaffold these systems toward these instances. Therefore, whilst we frequently carry around smart devices to view other people's digital representations (often through representations in social media) online, the ability to use these devices to support social interaction with those people in face-to-face encounters is under-utilised [38]. Part of this endeavour entails considering how these digital representations of ourselves layer toward impression management both in motivation and in impression construction [69].

The digital representation technologies are predominantly focused around specific devices rather than across devices, such as smartphones [20–22], smartwatches [32], and head-mounted displays [31, 46]. Beyond the context specifics mentioned above, there are constraints involving the technology affordances of these devices, from users sharing a device's screen and using the screen as common ground to support the conversation with smartphones [20–22] to sneakily utilising a head-mounted display to view other users [31]. Nonetheless, current investigations are confined to a single technology instance, with users often in a 'one-on-one'-type scenario wherein all people have the same technology at their disposal. There is further motivation, then, for queries into device specifics when one wishes to build on such established devices for collocated social interactions as smartwatches [4, 7, 61] and smartphones [7, 55] toward head-mounted displays [7, 31, 46], which are only really starting to enter use in normal occurrences.

To build from current work, here we begin to investigate the impact of multiple device types for supporting social interactions when used together in a group setting, much as seen in everyday instances. In these settings, people use tablets, smartphones, laptops, smartwatches, and forward-directed head-mounted displays on social occasions to support behaviours – for example, showing a picture or supporting story-telling [59]. In creating social devices for face-to-face conversations, part of devising the devices' affordances accordingly, and part of our enquiry, involves looking at how to map these different behaviours toward supporting the various user interactions. Unpacking this question further, we find it important to study not only how people would interact with such devices together but also the perceived affordances and perception of this use. Without such investigations, the situated knowledge, as is the case now within this area of technology application, becomes device-dependent.

Holding two events with 28 users, we conducted an experimental study to explore using face-to-face digital representations within head-mounted displays, smartwatches, and smartphones to investigate digitally augmented co-located social interactions. This involved building on previously researched and tested software that presents users with digital profiles of other users in their location [32]. To triangulate our findings, we assessed our participants' interactions through group interviews, questionnaire analysis, video analysis, and measurement of device usage. The distinctions found between the technological interfaces can be used for future technologies in this field, to build a classification of device affordances. The implications of our findings are far-reaching for social-media designers and researchers who are concerned with how technology devices can support user behaviours and those dealing with face-to-face social-interaction technology.

## 2 RELATED WORK

Assisting collocated people 'to get to know each other' via technological solutions has been an active research topic since the classic work by McCarthy et al. [41]. However, less consideration has been paid toward the characteristics of devices aimed at providing this support. The evolution of devices that support socialising has progressed from large screens [41] such as tabletop devices [34, 44] to mobile devices, such as smartphones [20–22, 51], smartwatches [32], head-mounted displays [31, 46], and laptops [28] and from there to more experimental prototypes, including e-textiles [13, 24], coffee mugs [25], ambient displays [42, 56], and wristbands [8]. These different interfaces also play diverse roles within social interaction, from being a public display [20] where the screen is viewable by many people to acting as a private device [38] with its screen viewable by only one person. In addition, the devices' affordances vary between these settings, in distinct contexts – with smartphones, for instance, being potentially viewable by others [20]. The devices also go through paradigm shifts, as has been noted for tabletop devices, from lab prototypes to real-world collaborative applications and from single-touch to multitouch with multiple users [44]. The data held on and presented by said devices are often shown in user profiles. These representations are either mined from the participants' current online sources, such as Facebook [22, 47] and Twitter [22], or created for the context by the users themselves [31, 43].

### 2.1 Digital Collocated Representations

Aside from bespoke, novel prototypes, which are often developed yet have gone unstudied over multiple contexts [13, 22, 24], three types of devices have been examined the most in augmenting collocated interaction: smartphones [20–22, 51], smartwatches [32], and head-mounted displays [31, 32, 46]. As smartphones are ubiquitous, they are the most obvious device to use for studying how to support collocated face-to-face interactions. However, given the many everyday encounters that may benefit from digital augmentation [37], a smartphone may not always be the best delivery mechanism. For instance, a person would need to explicitly remove the device from a pocket or bag to access another person's information. When considering the variety of people and interactions that might occur, this is impractical and may significantly disrupt face-to-face interaction [20]. To reduce the effort required of the user, Jarusriboonchai et al. [22] have used proxemics – interpretation of spatial relationships, including study of the human use of space and its effect on behaviour, communication, and social interaction. Proxemics uses these factors to look at the social norms and to interpret and reflect upon the interaction in relative distance between people [16]. Greenberg et al. [16] have defined proxemics zones between intimate (less than 45 cm), through personal (45 cm – 1.2 metres), social (1.1–3.7 metres), and public (3.7–7.6 metres). Jarusriboonchai and colleagues [22], building on these definitions, used three varying proximity levels (2, 5, and 10 metres) to represent disclosure settings (user-created, Twitter, and Facebook, respectively). This system provides a way for the digital representation to reflect out-grouping behaviour through more 'private' data being available just when the users are within the same group face-to-face. In this preliminary work, Jarusriboonchai et al. [22] noted that users would glance at the screen of other people's smartphones when these were worn in badge-like fashion on a lanyard, changing its perceived affordances. In this fashion, research suggests that phone screens can be seen as public devices [22]. This method of wearing the display equally could have been applied in a less digital formulation (e.g., in paper prototype fashion or through drawings as employed by Peng et al. [53] with head-mounted displays); however, as paper forms have not been investigated in this connection, it is unclear what implications this loss of control might have for the interaction.

For providing faster and more ubiquitous access to information in these instances, smartwatches have been found to be more effortlessly glanced at, as viewing them only require the user to raise a hand [54]. Although the screen of a smartwatch is smaller than that of a smartphones, they have been found to display similar content [32]. Additionally, they can act as a public screen for neighbouring people – for instance, during meetings, when participants' hands often rest on the table [52]. This semi-public characteristic of a smartwatch, paralleling that of a smartphone, does raise challenges in the privacy domain, such as controlling who sees the display. Control over who is able to view it is a different issue with head-mounted displays, however, as the view is private, available for the wearer alone. Accordingly, Kytö and McGookin [32] found that smartwatches provided better support for grounding conversations, a process wherein the participants come to a mutual belief [3], than did head-mounted displays, because of this privacy feature. In prior instances, this grounding behaviour [3] has been represented by users pointing at screens, verbally referencing the content, and sharing the screen of smartwatches [32, 52]. In augmenting interactions with head-mounted displays, moments of miscommunication caused by the users having personal displays can arise, due to the common context being lost; grounding rectifies these [3], but this does not prevent the moments of lost communication when they occur [32].

Whilst this affordance of sharing a screen does not exist in the case of head-mounted display devices, they provide 'heads-up' interaction and can deliver information close to the line of sight [31, 46]. They hence are unlike smartwatches and smartphones, which are elsewhere, such as on the wrist, in a pocket, or held in our hands. Additionally, if the information is shown in small portions and close to the line of sight, the user can view the information without the other person noticing this use during face-to-face interaction [48]. That is not to say, though, that the divided attention is not perceived as weird and noticed by other users [29], and the occluding frames and optics covering the eyes have indeed been found distracting in face-to-face interactions [39]. Thus, the problem noted above with regard to a smartphone disrupting the interaction through its usage is present also with head-mounted displays, through a loss of eye contact [31, 39]. Also, whereas the challenges a shared screen creates with regard to divided attention do not exist for devices with head-mounted displays, these devices present the additional barrier related to shared attention instead. This sharing of attention has been shown to disrupt conversations, although in head-mounted displays the additional information can be displayed closer to line of sight than in use of a mobile phone, computer, or personal display [37, 50].

Commenting upon head-mounted displays for collocated interactions, Kytö and McGookin [31] argued that browsing people outside one's conversation group to find potential conversations is one affordance and use case of head-mounted displays in a multiple-user setting (though the perceived impact of this browsing behaviour has yet to be investigated). This is due to the users' browsing behaviours not being publicly available: a user can break social taboos against looking for someone else to talk to while still in conversation.

As presented here, clear differences have been found between smartphones, smartwatches, and head-mounted displays and in their everyday use in 'live' social interactions. A problem exists in that the specific deployment of technology within the study instances changes the context of the interactions and, hence, the research findings. Furthermore, as we have highlighted above, the affordances attributed to these devices are restricted to cases in which the same device type is used by everyone, whereas a multitude of devices is present in the wild. This multiple-device landscape is due to the ecology – for example, there are inter-user differences in smartphone preferences. Yet, when interface and digital content developers and designers contemplate which device to choose to augment face-to-face interactions, there exists no work comparing simultaneous use of these diverse device affordances for conversations.

In collaborative scenarios, technology systems can increase communication and comprehension among multiple users [34] if the technology is aligned with normal conversation rubrics [67]. Furthermore, 47% of people often use technology during meetings and conversations [9]. Yet, in the augmentation of face-to-face interactions, current systems are often positioned for one-on-one use [22, 28] or a small number of users [32]. While everyday conversations at social events commonly involve only about four speakers [30], people engaging in these conversations, as demonstrated in prior literature [32], move between conversations, leaving and creating new groups. Whilst the conversations are often within small groups, these groups interchange through a larger pool [30] as users browse the crowd [31, 32]. This pattern is not reflected in prior investigations of one-on-one multiple-user technology instances. Hence, a research gap is evident in social technology modelling for multiple-user conversations. Accordingly, this is our first research question:

**RQ1:** What are the device affordances in using head-mounted displays, smartphones, and smartwatches, and how do people use these devices in social collocated multiple-user interactions?

## 2.2 User Experience

Part of the dialogue around augmenting social interactions is looking beyond how people behave with these devices, to consider also the users' and other people's experience of this use (by self and others). Whilst 47% of people use devices in conversations and meetings, 51% of people find that behaviour inappropriate [9]. This points to the question of the impact of these social devices that we use to augment conversations and how we can avoid creating inappropriate experiences. Some of the effort to address these inappropriate experiences can be directed toward modelling how people wish to be represented, and towards how the representations affect our conversations. When presenting his theory of presentation of self in day-to-day life, Goffman [14] noted that people try to control and guide others' impressions of them, to avoid embarrassment of self and others. The personal boundaries for social interactions are managed online within and between social technologies and networks [12]. Recently, researchers have begun looking also at how people wish to be represented digitally for face-to-face conversations [22, 31, 32] and how digital representations derived from social media affect these 'real-world' conversations [1].

With regard to user experience, smartphones have been posited to be an ideal platform to augment face-to-face conversations. Indeed, users have indicated that hand-held objects are often seen as being an extension of themselves [34]. Such results suggest that having hand-held technology at one's disposal may aid in providing a connection between the physical and digital self. However, equally, the mere presence of a smartphone has been found to distract from social interaction, even when the device is turned off [55]. The lower social acceptability of a particular technology is evident also with head-mounted displays, because the user's intentions are totally opaque [9] and other people do not know what the device is being used for [29]. Acceptability does increase if the other individuals know the purpose of the device's use [29], but this disclosure conflicts with the user-cited benefits of personal devices, principally their function as a privately viewable screen [31]. Head-mounted displays are suspected to lead to misspeaking, overlapping (two or more people talking at once), and movement behaviours wherein the user performs unusual actions such as staring into space robotically to take in information [9]. Importantly, loss of eye contact too has been found to occur in social interaction augmented with head-mounted displays, for both the user and the other participants [32].

Proceeding from this, scholars have identified the importance of the role of gestures within the user experience of using smartphones, head-mounted displays, and smartwatches [7]. Therefore, guidelines [26] and gesture studies [70] are prepared for measuring users' experience across devices, to build consistent gesture sets [47] such that users can transfer interaction models and gestures [7]. It is quite pertinent to our discussion that gestures (macro and micro) such as pointing play a role in non-verbal communication in interactions [2, 31, 32]. Often with smartphones and smartwatches, macro-scale gestures are used, among them pointing and turning the wrist [10]. Pointing gestures include eye pointing [9, 29], gaze pointing, and hand pointing [2]. The gesture of sharing one's device screen to thereby ground the interaction [22] plays a part in the users' experiences of the interaction through communicating spatial information [2]. Sociocentric referential gestures of this type are practised and designed in such a way that these actions themselves provide additional conversation dynamics to allow further references within the conversation [18]. That said, these gestures frequently look socially awkward, can cause fatigue, and often lack precision since they do not always have a haptic surface – as is the case with head-mounted displays and other augmented-reality interfaces [10]. Addressing these cases, users employ microgestures, small gestures performed by the hands or fingers [71] and detected on a fine scale [10]. Often, microgestures are used alongside macrogestures in mixed-scale gesture interaction interplay that provides for generalised interactions, and then these get fine tuned, especially in bimanual interactions [10]. Alongside the gestures, portability and micro mobility (i.e., how a device can be manipulated and mobilised for various purposes [35]) plays a role in the dynamics surrounding device usage.

Accordingly, the role of gestures and portability is a vital element within technologies used for social interactions, where gestures inherently form part of the social fabric of a conversation already. Whilst gestures have been explored for smartwatches, smartphones, and a mix of the two [32], literature on how these gestures are used for social technology has been lacking, especially with regard to people's user experience for day-to-day social interactions across head-mounted displays, smartphones, and smartwatches in digital representation technologies [7].

Accordingly, as noted above, research into augmenting conversations has investigated people's perceptions of these social devices in individual instances wherein all users have identical devices – a scenario far removed from typical everyday occurrences. The variable of device difference and its impact, beyond speculation, has not been investigated to shed light on the role it plays in users' perceptions. Hence, our second research question is as follows:

> **RQ2:** What are the users' experiences of using various devices (head-mounted displays, smartphones, and smartwatches) in social collocated multiple-user interactions?

## 3 STUDY DESIGN

With the background discussed above, we sought primarily to investigate how devices and their affordances are utilised (RQ1) and perceived (RQ2) in multiple-user and multiple-device context. The investigation was done through a study wherein participants created a static profile and then were given a device (a smartphone, a smartwatch, or a head-mounted display) to use to view each other's profiles during a social event. This method built on the method Kytö and McGookin [31, 32] employed with smartwatches and head-mounted displays and Nguyen et al.'s [46] method for head-mounted displays. Here, we present the first work in this research space to look at this number of users and consider multiple technologies.

While work in this area has been done on support for networking [6], largely professional networking [46], our study setting is situated as a social event. Although people were invited to network, this was not instrumental exchange of contact details but interaction in a social and not-so-serious manner. The choice of devices represented the most popular ones employed for augmenting social interaction, typically within social situations, and excluded devices used in earlier work that were unavailable for comparative study [20–22, 31, 46, 48, 51]. In our research, two events were held, each 65 minutes long and involving 15 users, with five per device condition (head-mounted displays, smartwatches, and smartphones). The method was approved as ethical by Aalto University ethics board, with informed consent provided by all participants.

### 3.1 Procedure

Participants were recruited throughout the university and via social media, with a reward of two cinema tickets for their participation (2 × 15 euros in value). Before the event, each participant filled in a standard form providing demographic data, completed a relational assessment questionnaires, provided a profile picture, and created a static profile specifically for the event (see Figure 2). We choose a static profile because this approach is employed in current literature [22, 31, 32, 37].

### 3.2 Participant Relationships and Context

In line with prior work, we used a relations questionnaire, asking our participants whether they knew any of the other people taking part, to model the participants' social relationships [23, 31, 32]. In previous studies [6, 12, 19–22, 24] participants were required to be strangers (defined as people who have not already met). Here, the decision was taken instead to model the study as a more typical social event: participants usually represent a mixture of relationships. We classified our subjects into pre-defined categories, relative to each other: friends, strangers, and people known to the participant. A person in the final category was someone the participant had seen before but did not know, and 'stranger' denoted someone the participant did not know at all [cf. 31, 32]. We used these categories defined by Jones and O'Neill [23] because our study premise was similarly defined in terms of information for a social context on a digital device. These classes have proved useful to sort people into pre-defined categories, given that information-sharing varies with these relationship ties on account of contextual dynamics of sharing information [23]. Thereby, we designed the participant group in our study to be as normalised as possible and were able to measure the associated variables.

Participants were provided with context behind the social interaction, as context has been found to influence willingness to engage in social interactions [39]. The context given was '*an event to meet friends and socialise*'. To support an informal context, subjects were told that food and drinks would be provided at the event. This was to make the place and event informal, with a relaxed atmosphere avoiding the feel of artificial laboratory studies while the controls of a laboratory setting were retained as much as possible. We left the task deliberately vague, since interaction that lacks a clear *a priori* purpose is one of the most common situations of encountering people in day-to-day life wherein we predict such technology will be employed [37, 65].

### 3.3 Creating Digital Profiles

Using PowerPoint, our participants made image-based profiles for themselves that they felt represented them (see Figure 1). This method has been shown to work in similar settings [31, 32]. When creating their profiles, participants had been told of the context (as described above) and were also aware that the image would be presented on a smartwatch, smartphone, or head-mounted display to other users. We asked participants to give preference to images over text, since prior work indicates that participants find images to be better at supporting conversation [32]. A guideline of six images was given, for avoiding the ambiguity otherwise identified in previous cases of digital profiles [32] and for viewability on the various devices.

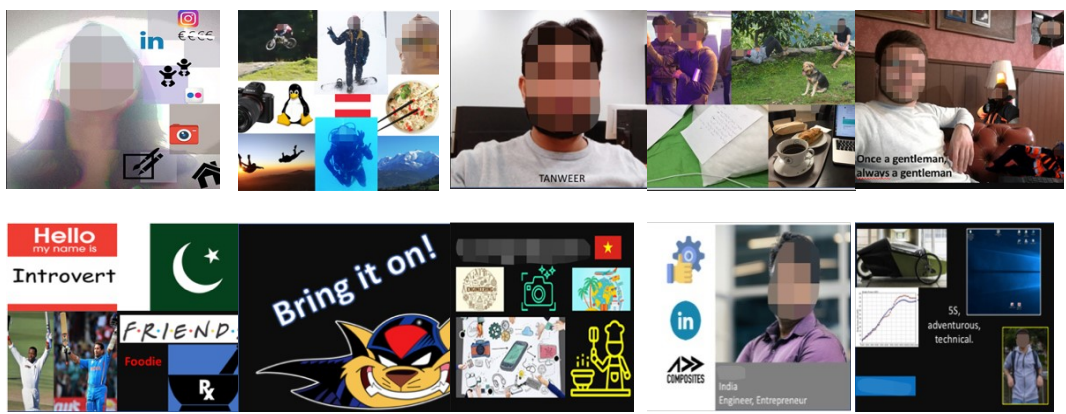

Fig. 1: A selection from the digital profiles created by participants.

We supplied a template along with our guidance, wherein the digital profile has a black background, to support with the head-mounted display system better. Except with regard to this dark background, the participants were free to design their profile in any way they chose, using as much or little media content as they wished. Participants were not time-limited in creating the profile. Some typical profiles can be seen in the figure. Because of scaling issues related to differences in screen size, we sent the participants images of their profile as displayed by the three devices (see Figure 2b). They could then change their profile and face image if they so desired. One participant did so because of not being able to see the profile clearly on a smartwatch screen.

### 3.4 System Mechanics

To support the viewing of profiles in context, we developed an Android application to work across the three devices. This was designed to display similar-looking content over all the user interfaces (see Figure 2). The devices we used were a Nexus LG 5Xd smartphone, an Epson Moverio BT-200 head-mounted display system, and a Sony SmartWatch 3. These devices were used to reflect common availability and what has been investigated for face-to-face interactions in prior work [31, 32]. The head-mounted system used was an optical device as this allowed the participant to maintain eye contact, which has been demonstrated to be an important factor when content is shown to a person in active conversation [31, 32, 46]. As Dingler et al. [7] have noted, each of these devices poses its own challenges to maintaining consistent interactivity. The software limitations of all the devices have been explored, and studies have shown specifically that they exist across different settings, in various contexts, and across different relationships. To address these, we made sure the devices were all similar in layout (user interface), to allow for information to be displayed similarly across all users while allowing swipe interactions in the smartwatch and smartphone conditions and permitting both use of button interfaces and interactions with the head-mounted display.

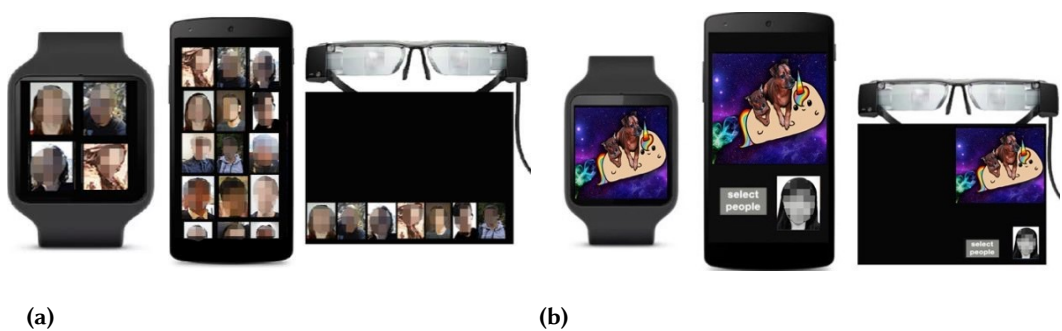

(a)          (b)

Fig. 2: (a) A smartwatch, smartphone, and head-mounted display showing the browsing screen where participants can view other participants' profile photos. (b) A user's profile presented on a smartwatch, on a smartphone, and on a head-mounted display. Note that the head-mounted displays' background is black, as this provides maximum transparency in viewing through see-through displays.

To facilitate configuration and support consistent timestamps in logging on a central server, the application timestamped entries indicating which profile a participant was looking at and how long the application was open.

When the participants opened our application, it presented a set of images of faces of other participants, which the user could scroll through (Figure 2a). By selecting one of the faces (either by tapping on the smartphone or watch screen or by using the navigation buttons on the handheld attachment accompanying the head-mounted display), the user would see the profile created by the respective individual (Figure 2b). Because of the limited space, participants could see only one profile at a time. The profile photo was next to it to act as a reminder of whose profile was being presented, except in the case of the smartwatches – since the screen was considerably smaller, a profile picture was not displayed here. In head-mounted displays, the users profile was toward the right for the user so as not to interfere with the participants' line of sight [32]. The participant could then swipe back, or click the Back button for the head-mounted display, to return to the browsing screen (Figure 2a). The list of other users in the latter view was static (they could scroll down to view more), as there were only 15 participants in all, making it easy to view all attendees on all devices. Were we to repeat this study with a larger number of people, this ordering would need to be done via proximity. The application displayed all people taking part in the study, not just those in the room at the time.

### 3.5 The Study Setting

A few days after creating their profile, participants were invited to attend one of two events. The study space (Figure 3), which was not on university premises, was a 25-square-metre room containing sofas, chairs, tables, and a buffet table with food and drinks. The light within the space was kept slightly dimmed to allow better use of the head-mounted display devices as the display is viewed better in these lighting conditions [64]. Three GoPro cameras were used to record the participants in the study space, one at the centre of the ceiling (see Figure 3a) and two in opposite corners of the room (the placement is shown in red in Figure 3b).

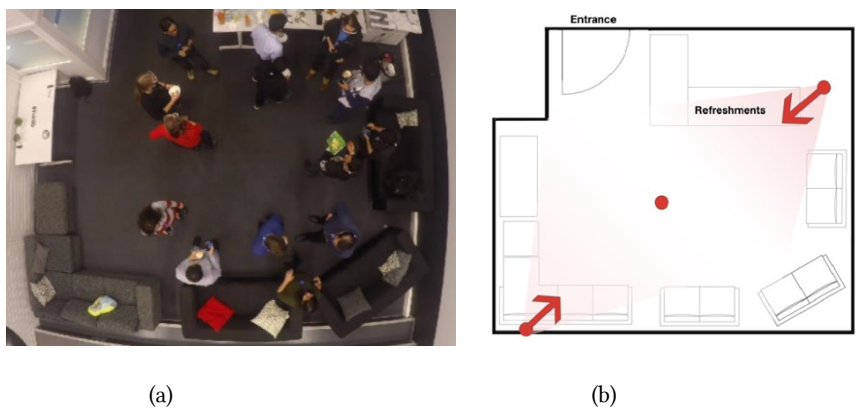

(a)          (b)

Fig. 3: (a) A ceiling-camera view of the study space (event 1 is shown here). The food and drinks provided are on the upper right, and the entrance is at top left. (b) A technical sketch showing the same room and its camera layout. The red arrows and circles represent the cameras and their filming direction.

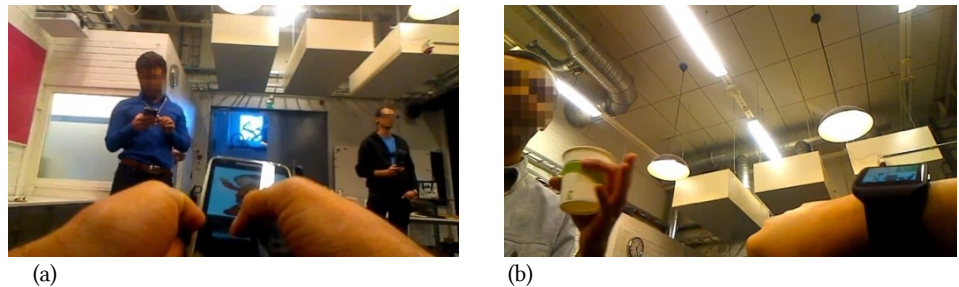

(a)      (b)

Fig. 4. Screen captures from the videos recorded by the wearable cameras: (a) from a participant with a smartphone, (b) from a participant with a smartwatch.

The participants entered the study space after arriving and stating that they were comfortable using the device. Participants entered in a staggered manner and in no set order. Hence, their order of arrival determined their order of entering the experiment. This method was designed to mirror a social gathering context, in which it is unusual for people to enter all at once. Also, the approach of having subjects enter the study room one at a time after being oriented to the study was chosen to allow us as researchers to introduce the technology properly to our participants, to allow them to maintain their relationship dynamics with strangers, and to aid in investigating the grouping dynamics. After orientation, participants were told that the study room is down the corridor, with the participants being free to enter that labelled room when they felt comfortable doing so. This also gave them a chance to use the technology to browse the profiles before entering from the corridor. No participant engaged in this behaviour, so it is not expanded upon here.

The orientation consisted of a researcher explaining the technology that the participant was going to use: a smartphone, head-mounted display, or smartwatch. The experimenter then helped the participant gain practice in using the device and browsing profiles (with 'mock-up' user profiles of celebrities). This ensured that the participant could use and access the technology whilst not giving prior access to the other users' profiles. Once the participant signalled being happy with using the device, he or she was equipped with a wearable camera (a small SnapCam Lite) and then directed toward the study space. The wearable camera, at chest height, recorded the participants' conversations and frequently also showed the wearer's and other participants' actions involving the devices (see Figure 4).

The study ran for 65 minutes, which came to 35 minutes after the last person entered. This gave all participants enough time to settle into the study, per prior research guidelines [30]. One key benefit of the method we chose is that it enabled participants to enter the study room when they felt ready and were able to use the device as in normal contexts, all without disclosing potentially confounding information before the subject entered the room. Also, each participant had access to the same digital content, displayed in similar form, allowing for information symmetry.

When the study was stopped, after 65 minutes, participants were given the relations questionnaires and a questionnaire on the event. The event questionnaire was composed of 41 questions: questions with a 1–7 Likert scale ('strongly disagree' to 'strongly agree'), yes/no questions, and multiple-choice questions about the devices. The questions covered the subject's behaviour (RQ1), perceptions as to other participants (RQ2), and technology preferences. Also, there were open-ended items asking about the participant's device use (RQ1 and RQ2) and future ideation. Lastly, for triangulation of the data, a semi-structured group interview was held with the users of each device. The interview questions focused on feelings about the event overall, technology aspects, the conversations, the users' profiles, social acceptability, and general feedback.

Whilst we aimed for a study model achieving an everyday, relaxed atmosphere, we recognise that this could not be the same as a normal event. Contributing factors are the inclusion of numerous devices, specifically head-mounted displays; the multitude of both wearable and fixed cameras in the study space; and the nature of the setting, involving a study. Though we made design decisions for the interaction to be as normal as possible, the study is still within the context of a particular locality, and the event had to be framed and orchestrated to such an extent that the findings possess necessarily limited generalisability. Additionally, the method described above is angled toward investigating the data with regard to our research questions' lens in technological mannerism. Further work could be done to take a social science stance to our data (e.g., by applying ethnological approaches).

## 4 PARTICIPANTS

The study involved two events, with 14 people each (for each event, one person did not show up), for a total of 10 users with smartphones (participants 6–10 and 21–25), nine users with smartwatches (participants 1–3, 5, and 16–20), and nine users with a head-mounted display (participants 12–15 and 26–30). As the participant numbers differ between groups, the data were normalised to be representative. Our participants consisted of 20 males and eight females ($M = 27$, range 22–39). Most users were strangers (91%), with 7% being friends and 2% classified as 'known people' (familiar strangers). Hence, our results are drawn toward strangers' collocated interactions.

## 5 DATA ANALYSIS

During the event, server data were collected from the application, which logged each participant's use of our system. The server data allowed for visually representing whose profile each participant was looking at and for how long (Figure 5 presents this graphical representation). This server data were used also for details of viewing times and overall usage, reported here as percentages and amounts of time. In addition, by correlating the various cameras in the study space (Figure 3) and the wearable cameras (Figure 4) with the server data (Figure 5), we encoded what a participant was doing digitally (from the server logs) with his or her physical behaviour (from the video material) to create an overall picture. This allowed coding for behaviours noted in prior work (grounding and sharing screens) and for behaviours newly identified via this process of correlating all the data gathered (the final codes are presented in Table 1). Thus, using the server data and verifying the timings through the captured videos enabled us to calculate usage statistics specific to the devices and participants. Prior work has demonstrated that this combination method for coding from videos and data logging is a successful approach for study of applications [31, 54].

The process of coding behaviours from the video material was developed via thematic analysis allowing the two lead authors to group the data thematically across the behaviours detected. This process, informed by the literature, followed typical video coding methods as have been previously employed with similar devices [60].

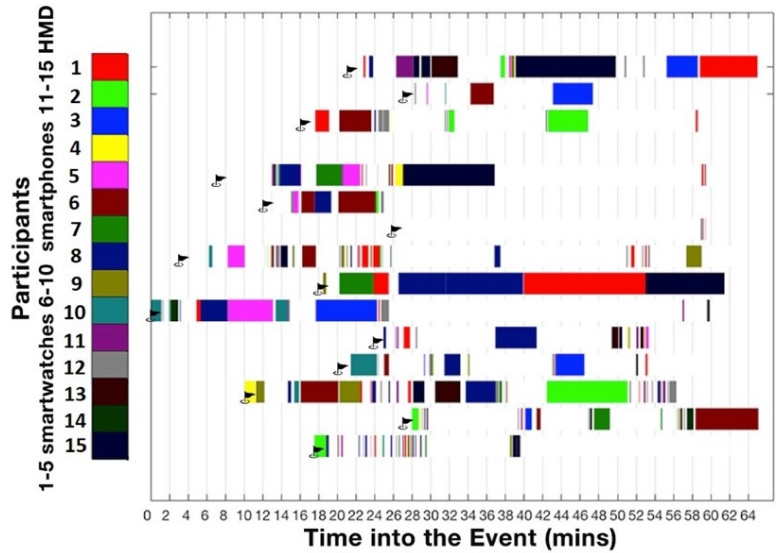

Fig. 5: Application log for event 1 for our 14 participants (P1–P15; P4 did not attend). Each user has a colour representation, which is plotted against event time elapsed (e.g., P1 viewed the digital self at the end of the event). The white portions of the timeline are times when the participant had no digital self element open. A flag symbol denotes when the participant entered the study event.

Whilst we did not opt for formal guidelines for the coding, we were aware of known behaviours, such as grounding, and built from these instances. Firstly, we devised codes from watching video footage of the recorded event, then discussed these and subsequently re-coded to refine the categories. This resulted in an initial coding schema along with the behaviour definitions. We then grouped these codes into themes around their use, such as sharing, conversations, and grounding. The results were further verified through three researchers then coding a 15-minute video segment from one of the events for all 14 participants independently. We calculated the inter-rater reliability of this coding schema to be 76% agreement. One behaviour in our labelling system involved terminology for what occurs inside and outside the conversation. We defined conversation-internality here as viewing details of a participant with whom one was engaged in conversation, whether an active speaker or a listeners (see Table 1). This was the primary source of disagreement within the codes: the timing of the behaviours relative to the participants leaving or joining a given group was not always clear (in some cases, there is no definitive point at which someone leaves/joins).

The final codes in Table 1 include behaviours previously identified with particular devices, such as grounding with smartwatches [31], showing a smartphone's screen [21], and browsing through users' profiles via head-mounted displays and smartwatches [49]. However, some of these behaviours were exhibited with other devices too, as with showing someone a smartwatch's screen, grounding via a smartphone, and browsing through users on a smartphone. The entirely new behaviours noted were handing over the device, seen with head-mounted displays and smartphones; participants viewing their own profile, across all devices; and three-way screen-sharing, seen with smartphones. These are later unpacked through their emerging codes.

Table 1: Behaviours noted during the study and their definition

| Behaviour | Definition |
| --- | --- |
| Browsing the users | A participant browsing through various profiles quickly, defined as a user looking at three or more profiles a minute |
| Viewing one's own profile | A participant looking at his or her own profile |
| Viewing those outside the conversation | A participant looking at users not taking part in the conversation being had |
| Viewing those in the conversation | A participant looking at users participating in the active conversation |
| Two-way screen-sharing | A participant sharing the device screen with another participant |
| Three-way screen-showing | Three participants all viewing the same device |
| Handing over the device | A participant giving the device to another participant to use |
| Grounding activity | A participant pointing or otherwise gesturing toward his or her own device or another person's display to reference the content (this is related to but distinguished from screen-showing) |
| Abandoning a device | A participant either removing the device (for a smartwatch or head-mounted display) or leaving the device away from the self (for a smartphone) |

Using this combination of data described above, the participant behaviours from the two events were then fully coded for each participant and event, including the study time elapsed. This coding was conducted by two researchers independently following each participant separately through the entire study. This produced, in total, 789 behaviour occurrences, averaging to 28 instances per person. After this, the final occurrence notes and end-of-study questionnaire were analysed via Kruskal–Wallis testing, since the data did not follow a normal distribution between groups ($W = .92$, $p < .05$). If significance was found between groups here, we, further, used ANOVA two-tailed statistical testing between groups to find the cause of the significant results.

Lastly, the group interviews were transcribed and encoded with ATLAS.ti by two researchers thematically from the initial video codes (presented in Table 1) to frame the work within this dimension, ultimately with the following interview codes: *viewing*, *browsing*, *grounding*, *sharing*, *abandoning*, *ignoring and distracting*, *group behaviour*, *social acceptability*, *behavioural*, *future use*, and *technological*. As earlier in the study, we were looking not at conversation dynamics but for insights with regard to RQ1 and RQ2 from the material. As for the video content, the codes generated went through a similar method of initially coding separately and then jointly discussing the codes to verify them and further develop definitions together.

## 6 RESULTS

The results are addressed in summary across the two events in accordance with the themes and codes noted through the data analysis. In general, participants rated the study itself as feeling normal ($M = 6.0$, 'mostly agree'), with smartphone users ($M = 5.5$) and head-mounted display users ($M = 5.4$) finding it somewhat normal and smartphone users ($M = 6.2$) agreeing somewhat that it felt normal. Contributing factors might include the inherent nature of the study setting and use of a new device and/or technology. On average, participants had no strong feelings about whether the technology influenced their behaviour independently of the devices ($M = 4.25$, 'neither agree nor disagree').

Considering the others present, participants felt that those with smartwatches appeared to act normally most often (82%), followed by smartphone-using participants (79%). Head-mounted displays' users were seen as acting abnormally (36%). This might be due to smartphones and smartwatches being present in everyday social situations while head-mounted displays are much less commonplace. All participants had used a smartphone device before, 25% of them had used a smartwatch, and half of the participants (50%) had used a device with a head-mounted display. Participants had a positive attitude to smartwatches ($M$ = 5.2) and smartphones ($M$ = 5.0) while exhibiting a neutral attitude toward head-mounted displays ($M$ = 4.5) ('neither agree nor disagree': 4; 'somewhat agree': 5). Most participants somewhat agreed that they liked the technology they were given ($M$ = 5.2).

## 6.1 Grouping

As there was a large number of people in the study space, participants would change groups often, forming and re-forming groups and conversations. At the events, participants joined and left a group, on average, between three and five times. The breakdown by device type was 3.2 group-joinings for head-mounted displays, 4.1 for smartphones, and 4.8 for smartwatches, on average. From the coding process, we noted that most of the group changes occurred when a user would go to fetch more refreshments.

## 6.2 Looking at Profiles: Viewing, Browsing, and Application Use

The technology provided a way to view and browse other participants and for the participants to view their own profiles (see Table 2 and Figure 2). However, participants found smartphones significantly better for this than smartwatches and head-mounted displays ($H(2)$ = 10.6051, $p$ = .005; head-mounted display $M$ = 4.3 (neither disagreeing nor agreeing as to ease of viewing), smartphone $M$ = 6.3 ('mostly agree'), and smartwatch $M$ = 4.2 ('neither agree nor disagree')).

Typically, participants browsed other participants' details when they first entered the study space, in rapid succession. This is visible in Figure 5 through the barcode-like appearances of multiple people's profiles after new people are shown entering. When we asked participants about this behaviour, they stated, for example: '*At the start, I just looked at all those profiles*' (P13, head-mounted-display participant). This browsing behaviour was often conducted to find someone to talk to, or whom the participant would be interested in talking to, in a pre-matching type of behaviour: '*There are actually people that I would like to talk to*' (P26, head-mounted-display participant). Participants often disclosed this behaviour to each other, '*It turns out that the person approached me because of my profile*' (P3, smartwatch participant), finding a ticket-to-talk in commenting upon things in the profile. This behaviour has been noted before [21, 49]. From the video coding process and examination of the data, we noted that, whilst almost all users began by exhibiting browsing behaviours right after walking into the room, not everyone browsed a person's profile before talking to him or her; e.g., one participant noted: '*I talked with some people without seeing their profile*' (P7, smartphone participant).

Table 2: Participants' viewing of profiles and time using the application (the frequencies shown are averages over all the participants)

|  | Total duration of profile views (seconds) | Total app usage time (seconds) | No. of profile views | No. of times viewing one's own profile |
|---|---|---|---|---|
| Head-mounted displays | 32 | 418 | 12.8 | 0.4 |
| Smartphones | 55 | 479 | 8.6 | 1.7 |
| Smartwatches | 55 | 440 | 7.9 | 1.3 |

Participants found it socially acceptable to look at the other person's profile ($M$ = 5.8, 'somewhat agree') and agreed that it is acceptable for others to look at theirs ($M$ = 6.2, 'mostly agree'). Since they felt this behaviour to be socially acceptable, users disagreed with the statement that they tried to conceal that they were looking at others' information ($M$ = 2.0, 'mostly disagree'). This is one of the results that could have been affected by the study setting altering the context of the interaction and render certain behaviours normalised.

In the interviews, some users of a head-mounted display stated that, because the device was private, they felt that they should ask the other participant for permission to look at his or her profile before viewing it and should disclose doing so: '*[I]f I did that with these glasses, I would feel socially compelled to ask the person beforehand*' (P15, head-mounted-display participant) and '*[T]his just makes it a lot creepier and, I think, a lot ruder than using a smartphone, where you at least see that someone is doing something*' (P27, head-mounted-display participant). However, it was not just users of these ultra-private devices who felt disturbed by their behaviour, with one smartphone-user, P5, commenting that '*I found myself very creepy. I saw someone entering the room, and I immediately was checking them out, and I felt creeped out by my behaviour*'.

Investigating this further, we asked participants whether they were aware of people looking at their profile. They found it most evident with smartphones (75%) and smartwatches (61%), with participants rarely noticing people viewing their profile with head-mounted displays (36%). This was noted by both the person whose profile was being viewed and the person viewing it, as with the comment that '*they didn't know that we were looking at the profile*' (P12, head-mounted-display participant). Conversely a few participants (11%) were not aware when someone was looking at their profile at any point in the study. Nevertheless, participants did not mind what device others used to view their information ($H(2)$ = 02329, $p$ = .890; $M$ = 4.9 for head-mounted displays, 5.2 for smartwatches, and 5.0 for smartphones).

When we delved further into the browsing behaviour, in a contrast to viewing profiles of others in general, participants did not always feel happy about someone pausing the conversation to browse other users within the application; for example, smartphone-user P8 said: '*I was talking to someone and they just suddenly started browsing, and I was just talking to no-one. I just felt very awkward.*'. As for how exhibiting of these browsing behaviours breaks down by device, on average, a user spent more time viewing a profile when using a smartphone or smartwatch than when using a head-mounted display (see Table 2). At the same time, head-mounted-display users viewed more profiles, on average, than members of either of the other groups, as Table 2 shows. Looking more closely at the duration of profile-viewing, we found that it increased toward the end of the event, to a greater extent with smartwatches than with smartphones or head-mounted displays. The correlation coefficient between the total time the participants spent viewing profiles and how long they spent viewing each one was statistically significantly higher for smartwatches ($r$ = .37) than for head-mounted displays ($r$ = .16, $z(2)$ = 1.96, $p$ = .049) and smartphones ($r$ = .19, $z(2)$ = 2.22, $p$ = .027).

This gives an indication that the technologies' affordances influenced both the number of profiles viewed and how long they were viewed, independently of the overall timing spent using the device. This could be due to the affordances of a smartwatch, where, as our captured video interactions attest, users left the application open on their wrist. In contrast, smartphone-users would lock the phones for safely holding it or put the device in a pockets / on a table and users of head-mounted displays closed the application to remove the image overlay from their field of view. Nonetheless, when one examines browsing behaviour over the full course of the study, it is clear that participants often exhibited large amounts of browsing behaviour at the beginning of the study, settling down into longer profile views as time elapsed (see Figure 5).

We also looked at our participants feelings' about their and others' behaviour related to browsing prior to a conversation. Our participants marked that they neither agreed nor disagreed with the statement that they looked at people before starting a conversation ($M$ = 4.4, 'neither agree nor disagree'), and they indicated a preference for looking at other participants once conversation had started ($M$ = 5, 'somewhat agree'), as with '*only looked at the pictures [profiles] when I was actually speaking with the person*' (P22, smartphone participant). In another interesting pattern, 35% of participants looked at their own profile first upon entering the room (35% of users with smartwatches, 50% of those with smartphones, and 11% for head-mounted displays). From the video analysis, we detected a high percentage of smartphone-users viewing themselves first, often connected with participants sharing their screen during an introduction to show another participant their profile in a verification mannerism and to support the conversation narrative as an ice-breaker (e.g., via a photo of the user in a particular activity). This highlights the importance of these social devices for not only browsing other users but also viewing one's own profile. Such openings and verification alike occurred with smartwatch-using participants also: '*[S]omeone who had a smartwatch showed me my profile on the smartwatch and he talked about it*' (P6, smartphone participant), and '*I showed my watch face to other people when I was asking whether that was them or not*' (P18, smartwatch participant).

The participants' viewing behaviour was coded also for viewing both inside and outside a conversation (see Table 3). This breakdown allowed further investigation of the affordance of the 'sneak viewing' possible with head-mounted displays, a behaviour also noted previously [29] in which participants view those outside the conversation to find someone else to talk with.

Table 3: Participants viewing those inside and outside their conversation (frequencies are given as averages over all participants)

|  | No. of times participants viewed people outside the conversation | No. of times participants viewed those involved in the conversation |
| --- | --- | --- |
| Head-mounted displays | 3.07 | 4.06 |
| Smartphones | 2.04 | 6.07 |
| Smartwatches | 2.00 | 4.04 |

We found that participants, overall, tended to view those in the current conversation twice as much as those external to the conversation group (see Table 3). Some stated that they would often '*try to go for the conversation itself*' (P16, smartwatch participant), viewing those engaged in the conversation often '*just for verification, "Okay, you are this person"*' (P5, smartwatch participant), as mentioned above. Therefore, future social technology would need to allow for viewing both internal to and beyond the current conversation but be primarily scaffolded toward viewing those already within one's conversation.

### 6.3 Sharing and Creating Context through Gestures

Throughout the two events, users often performed gestures to ground and support the conversations as a form of collaborative action, often by pointing toward the device, showing its screen, viewing someone else's screen, or viewing screens jointly with others (see Table 4).

Table 4: Participants sharing their screens, giving others their devices, and grounding with devices, where the frequencies are averages over all participants and 'N/A' denotes an affordance that is impossible with the given device type

|  | Instances of two-way screen-showing | Instances of three-way screen-showing | No. of times participants handed over their device | No. of times participants grounded with their device |
| --- | --- | --- | --- | --- |
| Head-mounted displays | N/A | N/A | 1.04 | N/A |
| Smartphones | 5.2 | 0.04 | 3.05 | 7.03 |
| Smartwatches | 2.6 | 0 | 0 | 3.02 |

Grounding was frequent; as one subject stated, '*if I want to show you something, I will show it from my screen*' (P24, smartphone participant). Prior research has shown that grounding 'repairs' a conversation when moments of miscommunication occur [21, 32]. Whilst our participants somewhat disagreed that there were moments of miscommunication ($M = 3.6$), they acknowledged that miscommunication did occur. They indicated that this happened mainly with head-mounted-display users (32%) rather than smartphone-users (18%) or smartwatch-users (11%). As a head-mounted-display participant commented, '*Yes, I lost concentration. Like, while I'm sitting here if I'm using a phone it's not that prominent because it's part of daily life; with your eyes focusing somewhere else, it looks like a person is distracted*' (P29). Another in this device group, P14, stated: '*I was looking into someone and they thought I was talking to them, but I was talking to someone else just because I was checking on things.*' Thus miscommunication often arose with head-mounted-display participants since what they were viewing, where they were looking, and at whom was not publicly visible as is the case with smartwatches and smartphones. Additionally, these users could not ground by sharing their screen or by using referential gestures, such as pointing. These participants felt that they were at '*more of a disadvantage […] if I wanted to show another person on my device, I couldn't*' (P13). Grounding behaviour was seen mostly with smartphones (with an average of 7.03 instances per user) and not quite so often with smartwatches (on average, 3.02 instances per user) (see Table 4). Often, grounding behaviour was initiated by someone pointing toward the screen and then sharing it with another participant by way of verification (as mentioned above).

Though most participants viewed others' screens, either of smartphones (54%) or of smartwatches (46%), interestingly, some people did not view anyone else's screen at the event (21%). Whilst the average user of a smartwatch shared the screen only a few times, the behaviour took place significantly more when the sharer was using a smartphone, as is visible from Table 4 ($H(2) = 12,804$, $p = .002$).

Screen-sharing occurred beyond dyads also – i.e., in groups of three people – but only with smartphone devices (see Table 4). It is consistent with this that most participants shared their device with only one other participant (a behaviour exhibited by 56% of smartphone- and smartwatch-users). Users of head-mounted displays had to give someone else the device physically to share it (in fact, 78% of head-mounted-display participants actually did this). As the table shows, equivalent device hand-over did not occur with smartwatches. We suspect this is because the watches, while similar to a head-mounted display in being worn, cannot as easily be removed; head-mounted displays can be lifted off, while handing over a smartwatch requires the user to undo a strap. As for smartphones, we would expect hand-over behaviour to occur less among people using their personal devices (rather than study units); however, this is an area open for investigation.

This behaviour of publicly sharing the screen was not always self-initiated. Users sometimes glanced at each other's screens (with 61% of participants admitting to looking at others users' smartphones and 50% for others' smartwatches). A few participants did not look at someone else's device in this manner (19%). In general, our results indicate that smartphone and smartwatch devices' screens are seen as a public viewable space even though the device is private, whereas a head-mounted display is private in both respects unless physically handed over by its user.

### 6.4 Abandoning, Ignoring, and Distraction

While participants had been given the device, they did not always use it throughout the study. Instead, they often would find that the devices distracted from the conversation so abandoned them, whether by putting them in a pocket, leaving them on the table/sofa (smartphones), or positioning the head-mounted display unit on top of the head as with regular glasses. No participant removed a smartwatch, so these were abandoned significantly less than head-mounted display units and smartphones ($H(2) = 12,783$, p = .002; average number of times a participant abandoned a head-mounted display: 1, a smartphone: 3.5, a smartwatch: 0).

Head-mounted displays were said to be abandoned to reduce distraction – '*if you have the glasses, they're always on, it's always there in your vision, and so that's a little bit distracting*' (P27, head-mounted-display participant) and to appear more normal '*I would just prefer a smartwatch or mobile device, because it's one glance to look, and then I'm not holding it all the time and so it goes in the background*' (P14, head-mounted-display participant). That the requirement of having the device inhibited participants' normal behaviour was highlighted with smartphones also: *[T]he phone was the most annoying part [...] that it was in your hand and you couldn't really put it anywhere. So, it took one hand*' (P9, smartphone participant). It is worth noting that this could be an impact of the study design, wherein food and drinks were offered and cups and plates held in the hands, but is indicative with regard to normal occasions on which things are held in the hands during conversations [35].

Most participants agreed that they were listening to the people they were having conversations with when using the devices ($M = 6.4$ for smartphones, 5.8 for smartwatches, and 5.3 for head-mounted displays). It became apparent with head-mounted displays, however, that the focus of the user's eyes had an important part in the social interaction for both the user and the conversation partner: '*when I'm looking at something, they think that I lost concentration*', so '*I wanted to get rid of them [the glasses] at some point when there was no need for looking at someone's profile*' (P28, head-mounted-display participant). Because of device use, most people felt that often other participants were not fully listening (39%), mostly those using head-mounted displays (32%), followed by smartphones (18%) and smartwatch participants(11%); for example, smartwatch-user P20 said: '*I had a conversation with one of the glasses-wearers where they'll just stare off into the void, into nothing.*'

The distraction that the technology caused within the conversation was noticed primarily by people talking to someone wearing a head-mounted unit rather than the wearer him- or herself, where it was found difficult to ignore that the participant was wearing the device ($M = 3.4$, 'disagree'). This is in contrast to smartwatches, for which it was easy to ignore that someone was wearing the device ($M = 6.0$, 'mostly agree'). When the device was in active use, the conversation partner found it significantly easier to ignore that the other person was using a smartwatch than to ignore use of smartphones and head-mounted displays ($H(2) = 11.2318$, $p = .004$; $M = 4.7$ for smartphones, 5.6 for smartwatches, and 4.0 for head-mounted displays).

The ability to ignore the device being used resulted in participants finding head-mounted-display users to be significantly more distracted from conversations than were participants with smartwatches ($H(2)$ = 7.236, $p$ =a .0268; head-mounted displays vs. smartwatches: $t(18)$ = -3.18192, $p$ = .002) ($M$ = 5, 'somewhat agree'). This is consistent with wearing of smartwatches being found significantly easier to ignore than holding of a smartphone ($t(18)$ = 2.05339, $p$ = .007) or wearing a head-mounted unit ($t(17)$ = 2.58641, $p$ = .02] ($M$ = 3.1 for smartphones, 5.6 for smartwatches, and 3.4 for head-mounted displays).

## 6.5 Device Preferences

Overall, participants stated that they would have preferred to use head-mounted displays (50%), followed by smartwatches (35%), with the fewest people expressing a preference for smartphones (29%). As people could pick more than one device, 11% of people listed head-mounted displays and smartwatches, and 4% of people wanted smartphones and smartwatches. From looking into the preference data by separating out the primary device participants used, it is clear that the subjects, regardless of the device that they used, wanted head-mounted displays (see Table 5). Additionally, we noted that smartwatch- and smartphone-users both preferred their assigned device the least. All of our participants wanted to use some device in our study context (as Table 5 shows).

Table 5: Participants' preferred technology for viewing digital profiles in social conversations, by the device assigned (rounding is to full percentage points, and the figures sum to more than 100 because participants could pick more than one option)

|  | Technology preferred | | |
| --- | --- | --- | --- |
| **Technology used** | Head-mounted display | Smartphone | Smartwatch |
| Head-mounted display | **55%** | 33% | 33% |
| Smartphone | **50%** | 30% | 30% |
| Smartwatch | **44%** | **44%** | 22% |

The open-ended material let us dig deeper. Participants stated that not wanting smartphones stemmed from use of a phone containing social signals that the participant was '*bored of the conversation*' (P10, smartphone participant), but they did like that smartphones are '*easy to use*' (P23, smartphone participant). As for affordances, smartphones can be put in one's pocket, so participants can '*use it and you put it away*' (P24, smartphone participant) or abandon the device in such a way that '*it's there, but not exposed*' (P8, smartphone participant). Participants acknowledged that the smartphone screen is a publicly viewable device '*you are trying to see, and another person knows what you're doing*' (P25, smartphone participant). Participants pointed to the publicly viewable display of smartwatches and smartphones as something by which '*you can see*' (P2, smartwatch participant) what people are viewing, while this is harder with smartwatches since '*with the watch it's just like... I don't have such good vision*' (P18, smartwatch participant). Participants who wanted smartwatches said that they '*feel like it will be more comfortable*' (P8, smartphone participant) though '*it's not as big as on the mobile screen*' (P17, smartwatch participant), also noting that '*even if you are not using it, it's always hanging there*' (P3, smartwatch participant). Participants wanted the head-mounted-display devices '*just for the novelty factor*' (P21, smartphone participant), '*to try them*' (P1, smartwatch participant), as '*wearing glasses was* *much cooler than using the other devices*' (P15, head-mounted-display participant), and some pointed to the ability for the viewing behaviours to be '*a secret*' (P14, head-mounted-display participant). Nonetheless, these participants recognised that head-mounted displays are '*a complete distraction*' (P13, head-mounted-display participant) and that it is '*a bit uncomfortable as well, wearing it*' (P12, head-mounted-display participant). Those who had not used them still '*want to experience how the smart glasses work*' (P2, smartwatch participant).

The main reasons participants stated for wanting to switch technology included the assigned device being uncomfortable (head-mounted displays), wanting something wearable (smartwatches / head-mounted displays), and finding the technology distracting or awkward (smartphones / head-mounted displays). As a smartphone user noted, '*I would prefer a smartwatch, personally, because I don't need to carry this or wear this*' (P6). Thus, preferences for each device were in line with the requirements and affordances expected with each individual device type.

Unexpectedly, with regard to other people, participants preferred others using head-mounted displays (46%), followed by smartphones (38%) and smartwatches (32%). A small proportion of users preferred the others use no technology at all (4%). The preferences for others' use can be itemised by assigned technology thus: users of head-mounted displays were in the strongest favour of others using head-mounted display technology, and smartwatch-users preferred the other options, smartphones and smartwatches (see Table 6).

Table 6: Participant preferences for technology that others can use to view digital profiles in social conversations, by the device assigned (the figures are rounded to the nearest percentage point, and, because participants could pick more than one option, sum to more than 100)

| Technology used | Technology preferred for others' use | | | |
| --- | --- | --- | --- | --- |
| | Head-mounted display | Smartphone | Smartwatch | None |
| Head-mounted display | **55%** | 44% | 33% | 0% |
| Smartphone | **40%** | 30% | 20% | 10% |
| Smartwatch | 33% | **44%** | **44%** | 0% |

Participants who had not had a head-mounted unit as their assigned device thought it would be '*easy to use getting info*' (P3, smartwatch participant), '*flexible*' (P24, smartphone participant), '*innovative, therefore attractive*' (P21, smartphone participant), and '*cool*' (P25, smartphone participant) but were worried about the '*camera and recording*' (P6, smartphone participant). Head-mounted displays' users felt that devices viewable by both their user and others, such as smartwatches and smartphones, are better because '*I can see what they do WHEN they do that*' (P14, head-mounted-display participant). Head-mounted-display participants picked other devices because their assigned device had '*overlaid the real world all the time*' (P28, head-mounted-display participant) and the other devices '*feel more natural*' (P27, head-mounted-display participant). This draws attention to our finding that participants in this group wanted others, as well as themselves, to use these devices for social interactions. Our one participant who did not like our technology commented also on finding '*no need for looking at someone's information*', a remark on the whole concept of the study (P10, smartphone participant).

# 7 DISCUSSION

The findings from this study have use in two main areas; the first is in the study of device affordances with regard to supporting face-to-face augmented interactions (RQ1), and the second is in the exploration of how people experience the use of these augmentations with respect to their interactions (RQ2). Considering these implications leads us to certain user interface recommendations and design guidelines for social applications in face-to-face use to help frame this work within the larger body of knowledge.

## 7.1 RQ1: Device Affordances

From the results, it is evident that the device capabilities and affordances influenced our participants' behaviours. One outcome of our coding process was to reveal generalised stages of interaction. These were (1) *browsing*, (2) *verification*, (3) *viewing*, and (4) *settling*. In the first stage, *browsing*, a participant would browse through many users quickly, often in isolation. This can be seen in Figure 5. The participant would then join a conversation in progress or approach another user, often with that user's profile open or opening it shortly after approach, for *verification* purposes. This can be seen in Figure 5 as people looking at their own profile. With participants using head-mounted units, this verification instead was done verbally since the screen was unable to be shown. It is in this period that most grounding and showing behaviours with smartphones and smartwatches were exhibited. Once the conversation began, participants would *view* those involved in the conversation, especially with smartphones, and occasionally *view* those outside the conversation, particularly if using a head-mounted display. For head-mounted-display and smartphone participants, this is the time in which the device would be handed over either to ground the conversation or to let the other user 'experience' their device. After this introductory period, participants would leave the profiles open longer, often without attending to them, with some participants abandoning the device, in the period we refer to as *settling*. This can be seen in Figure 5 as the longer periods of viewing with fewer profiles being viewed. The behaviours identified at this stage, however, were dependent on the device used. Nonetheless, these four stages provide researchers and those interested in social augmentation devices an initial method for setting the stages of interaction in a framework. However, further work is needed to scaffold our concepts toward general everyday usage outside our experimental set-up.

Table 7: The devices' affordances for digital representation – how collocated technologies supported behaviours

|  | Screen-showing | | Grounding (pointing) | Handing over the device |
|---|---|---|---|---|
|  | Two-way | Three-way | | |
| Head-mounted displays |  |  |  | ✓ |
| Smartphones | ✓ | ✓ | ✓ | ✓ |
| Smartwatches |  |  | ✓ |  |

While these stages were generalised across all devices in our study for the most part, some device-specific profile-sharing behaviours and gestures were displayed. Generally, smartphones supported most of these whilst head-mounted displays supported the fewest (see Table 7). These affordances are influenced by device factors identified here as screen size, accessibility of sharing, viewing disclosure (public vs. private), and mobility.

Highlighted here, micro mobility / portability supports the user's communication by making possible certain actions, such as handing over and abandoning the device. This factor, along with a larger screen, allowed our participants to use microgestures and macrogestures across device type. For instance, smartphones, relative to smartwatches, were more easily handed over, and a user could manipulate the former freely without having to turn the wrist to show the screen and thereby create an unnatural interaction. Accordingly, smartphones have been shown to entail more natural interactions, as commented by our participants. This limitation of smartwatches has been noticed before as part of their modality, and in this case micromobility is relevant too, as is the small and usually poor display [61]. However, our results indicate that more work needs to be done to fully recognise the microgestures within the study instance through finer measurements.

Part of these gestures is the availability of information to be gestured toward to create mutual understanding (i.e., the grounding). Information in our study was either publicly displayed (on smartphones and smartwatches) or privately viewed (via head-mounted displays). Another relevant aspect of our exploration involved the social acceptability of these behaviours, with our participants noting in particular that they looked at each other's screens. As Koelle et al. [29] mentioned about head-mounted displays, a person's usage becomes more socially acceptable when the activity is known, and our findings are consistent with this. We also found support for Ens et al.'s [9] findings; it is evident that it is easier to ascertain the activity (in this case, what the user was viewing) when it is presented on a smartphone as opposed to a smartwatch. Jarusriboonchai et al.'s [22] work took advantage of these factors and also showed that use of a smartphone is more acceptable when the information is presented directly (in their case, as outward-facing badges). The issue is further compounded with head-mounted displays, with which it can be difficult to view the content against a non-uniform background [51], as one might expect in the case of conversations [27]. That said, the head-mounted displays we used were chosen for this very reason, since they allow for the easiest distinguishing between the real and virtual scene [27, 53].

It was evident in our study that in social technologies, sharable devices are seen as public displays (even in the case of private devices), especially those with larger, easily visible screens and that can be easily handed over (or set down etc.), also affording multiple gestures and high micromobility/portability. This highlights a tension between privacy and the ability to share in technology to aid social interaction, wherein one mitigates the other, aligning our findings with prior work [23, 31, 32]. We noted a phenomenon in which participants changed their viewing behaviour on the basis of the public viewability of the screen and, hence, did not need to control other people's perceptions of their viewing habits. For example, the participants with head-mounted displays viewed more people, in quicker succession, than did users of other devices, since they did not have to manage other people's impression of their usage as did those with public screens, of smartwatches and smartphones. Here, we must reiterate the overarching layer to our study in which the participants were using study devices (i.e., not their own devices) and were limited to using them to view each other. The activity was known and limited. We suspect that if the same software were deployed in a real-world instance, the social acceptability and usage would be different, as this takes it outside the restricted and safe confines of the study and into a context wherein half of people find technology use inappropriate [9].

Our results imply that social behaviours that govern our interactions with each other are also established within people's interactions involved in our device usage in face-to-face scenarios. As we noticed in our study, the affordances of the devices with regard to aiding in this extension of self shape our behaviour in our interactions. This pattern has been noted before in the setting of online social networks [12] and in face-to-face conversations involving technology [39]. Part of the affordance of social interaction is to afford smooth and appropriate-seeming division of attention between the different agencies. In our case, attention is directed toward other people, their digital profiles, and the technology during the interaction. In such situations, often the users between themselves apply joint attention wherein sharing is inter-subject meaning-making [66]. In our case, joint attention was noticed in direct gestures such as pointing to reference a certain aspect of the profile or verbally indicting that one was looking at a certain person, for our head-mounted display subjects. This allowed users to ground the conversation through having a shared focus.

We have alluded to certain behaviours that may assist in reaching joint attention through gestures, but we do not yet know what it would mean to fully quantify joint attention between the technology and its users, especially in generalised contexts. For instance, the moments of miscommunication at our events were often due to a failure in joint attention that stemmed from head-mounted-display participants misleading other people through their referential gazes. Additionally, head-mounted displays often distract from conversations, as has been found previously too [39]. Therefore, it is not so clear what might be quantified as collaborative in this scenario and as a group phenomenon. A future step in this field of this research, toward defining and quantifying joint attention aspects, could be to monitor and adapt to the users' behaviours, creating attention-aware systems [57]. These systems could adjust to the highly dynamic environment of the conversation, supporting multitasking of cognitive resources for the attention processes between using the technology systems and engaging in conversation.

As our literature review highlighted, the behaviour of private viewing has been identified before as a way of finding new people to talk to without the other users noticing [32]. However, in a contrast against what Kytö and McGookin [31, 32] found for their application, the concomitant inability to ground and share, coupled with head-mounted display participants often seeming to act strangely in their eye movements when undertaking this behaviour, was seen to hinder communication significantly. The differences could well be due to our work making available multiple devices and more people to browse. These facets of setting emphasise that there is more work to be done to understand the social impact of systems for face-to-face conversations when multiple device types are to be integrated into and between conversations. One aspect of our study that makes it valuable in this regard is that participants were able to offer insights by drawing from their experiences with other devices and the evident affordances in our context. This represents a clear advantage, in that the presence of multiple devices can make up for missing device affordances, allowing people to have grounded perspectives and reflect upon various technologies.

### 7.2 RQ2: User Experience

The device affordances influenced the behaviours that we identified and also the participants' perception of their own and others' use of our system. This was spoken about in terms of the participant appearing and acting normal. Users' references to normality were focused on factors that could lead a technology to distract from conversation or, equally, enable the participant to listen and be listened to. Technologies, such as head-mounted displays and smartwatches, which are technological augmentations of our normal wearable items (glasses and watches), were seen as allowing the user to augment their conversations whilst appearing commonplace as they are unobtrusive [4]. Still, there was a fascinating disparity between the preferred device and the versatility of affordances provided (see tables 4–6). Key examples are that, whilst head-mounted displays created the most moments of miscommunication and most people felt that users of these did not activity listen, head-mounted display devices were still the device favoured most by both that device type's users and other participants in the conversations. Thus a dilemma emerges, wherein the technology that allows the user to behave normally by being unobtrusive with regard to the conversation hinders exchanges via miscommunication that hampers sharing of context.

It is worth noting here that there is a wide range of familiarity with the devices tested. Even though 50% of our participants had used head-mounted displays previously while only 25% had used smartwatches, the former is the more unusual device, not used in typical everyday contexts. Accordingly, there may be emotional reservations as to its use in comparison to smartphones and smartwatches, which could bias the results toward the more familiar technology. Such novelty and unfamiliarity factors need further investigation for uncovering further nuances of the devices.

## 8 RECOMMENDATIONS FOR FACE-TO-FACE SOCIAL APPLICATIONS

With regard to closer scrutiny of how devices are perceived and their affordances for augmenting our social interactions, it is evident that the way these systems are used is a multifaceted articulation of diverse behaviours, social and conversation norms, engagement mechanisms, and other values. A simplistic approach considering only one technology does not acknowledge the nuances behind the complex rituals that weave people and systems together for day-to-day life. As Rogers [58] notes, technologies are need to excite and engage people, allowing them to do what they want to do, need to do, or have never even considered before. Drawing from our results, we find it evident that, in this interplay, what affordances a device can support and what a user requires are two distinct aspects to be overlaid on how to design and build applications for face-to-face interactions and in gathering user requirements for these communications. Our work is the starting point to an acumination of studies, where further work is needed to ground and generalise our findings with sociological methods. Nevertheless we provide initial findings from our experimental approach, which can inform technology recommendations.

As noted in our introduction, we predict that technologies such as the one presented here could be employed in day-to-day life. For this situation, we suggest that the use of smartphones is to be preferred on account of this being a ubiquitous technology and supporting multiple end-user behaviours. For system designers, this device offers the broadest range of user interactions on which to build future applications. However, if we take a more user-centric approach and proceed from what the end stakeholder desires, this runs counter to what would be more relevant for systems designers, design of head-mounted displays. Our findings highlight that further work needs to be done to discriminate among our participants' complex motivations and ascertain the applicability in day-to-day life and the likely reactions of those not aware of or part of the interaction. As Koelle et al. [29] note, head-mounted displays are much more acceptable when people know what they are being used for, as was the case at our events. In such settings, they gain in social acceptability. This gives further impetus, supported by our users, to pursuing the notion that head-mounted displays are good for the sorts of task in which all users are restricted to using these displays, with the collocated others having symmetric information.

One aspect continually drawing the attention of our users was the ability to engage in private viewing whilst in a public situation. This may point to an ideal situation for both parties: a smartphone that offers the augmented viewing aspects of head-mounted displays, with the screen viewable by only the user, free from social stigma, whilst equally possessing the affordance of shareability with multiple other users.

It is perhaps in light of this realisation, then, that we recommend that designers of face-to-face technologies mitigate between affordances and user requirements, taking the best facets found in the various technologies for the applicability of the approach. This is not to say that only one device can be used at a time; different devices could be used by the same user, at different stages in the interaction. While we often switch between devices in day-to-day life, how plausible such implementation could be or how normalised this behaviour is within this instance space is open for question, but users could begin by viewing people via a head-mounted display, during the *browsing* phase, then continue the interactions through smartphones and smartwatches used in *verification*, *viewing*, and *settling* over the course of conversation.

For the future design of technologies for the space around social applications, we would posit with regard to software development that information presented to users needs to be mitigated toward the easily accessible with a low cognitive load such that it does not interfere with the conversation and the ability to be hidden or put away remains. This is especially true about information in the line of sight. Our findings show the balance to be important as, whilst these technologies are aimed at initially creating an ice-breaker, the continual presence can distract and cause breakdowns in the conversation that are counter to the device's intentions. Therefore, we stress that future work should look further at this boundary between displaying information during conversations and maintaining normal conversation practices. Attention-aware applications may help users navigate this space.

We also advocate that the information shown to a user include that user's own information (and profile picture) for referencing purposes (*verification* here) and ensure that all participants are presented with the same information. As for the profile user interface, it should be as large as possible within the limits of the device, as users often include lots of information here. However, this image when presented by a head-mounted display should not interfere with the line of vision, to allow the users to maintain eye contact. As shown in our work, the issue can be mitigated against through verifying the profile with the other user, and we argue for this, but quite often small details still are included. One solution to this problem is to allow the users to put their devices together when in groups, thus forming larger displays as in the work of Lucero et al. [34] on photo sharing, or to allow the users to share information across screens in real time in a joint mannerism. For displays that are entirely personal, such as head-mounted devices', one possibility could be to implement sharing features across device types within a given conversation group to support this behaviour. The implementation could employ a few tap operations, sending a request to the receivers (i.e., the other participants), who can decide whether to view the screen offered or not. We suspect that this sharing application would also allow others within the conversation to feel more comfortable with what the display's user was doing, as indicated by Koelle et al. [29], and his or her eye movements as viewing actions and motivation would become explicit. As in Jarusriboonchai et al.'s [20] application, an outward-facing screen that displays what the person is looking at makes devices more socially acceptable. This shareability could also work toward intersubjective intentional joint attention to a shared meaning in itself. The shared space then become more than static, to further support the conversation.

Equally, we highlight that it would be interesting to enhance the development of what these profiles hold and would look like in building from the digital self [31, 32] and profile [22] approach in current socio-technologies. Part of looking at the user experience of our software (RQ2), even though this was not directly addressed in the paper, was the representations of self, what they hold, how they are curated, and how they can be used to shape the users' interactions. Although we maintained a relevantly small user group of mostly strangers, how these profiles might look for a world-wide public audience, friends, and family is interesting. Scholars could fruitfully investigate how the framework presented above would scaffold for supporting these conversations. Logically, with known people, the *verification* stage would not be required, so it would be of interest to see how to adapt face-to-face software for diverse relationship dynamics.

Our final consideration in the realm of software recommendations is to support all four stages of behaviours among strangers as typically engaged in: the entire *browsing*, *verification*, *viewing*, and *settling* cycle. For hardware recommendations, we would suggest technological systems that can be easily removed from the conversation and, accordingly, be scaffolded toward the specific conversation and the person's attention and affordances. We recommend using devices or augmentations that do not occupy a hand continually or require strange and unnatural eye movements that could distract from the conversation. Ideally, the technology should blend in with everyday conversation through augmenting a currently used device but not blend in so much that it cannot be removed altogether to allow the end user to scaffold the device to the conversation him- or herself, becoming socially seamless.

## 9 FUTURE DIRECTIONS AND LIMITATIONS

Whilst our study provides significant steps forward in the form of a wider evaluation drawing from the literature, here we have considered only three possible interactive devices, with none being ideal on its own for the task. Nevertheless, this work refines and improves upon prior investigations of face-to-face social interaction by aligning the islands of knowledge together to permit contextualising past instances. In addition, although our scenario effectively simulates many characteristics of a social gathering, by necessity we are still combing through initial impressions on account of the novelty of the interaction, not all participants having used all the devices previously, and the software design possibly having an influence. A future iteration of this study could include a control group to mitigate against the technology factor and allow our participants to co-design profiles with us.

As was highlighted in the introduction, equally our findings indicate that some behaviours appeared normal in our setting, such as looking at digital devices when one was in conversation, which may not be the case in real-world instances but connected with our study setting. This may mask that fact that, whilst social-media profiles are an established phenomenon, the implications of using software that presents others with information about the people around them, as done in our study, are not fully known with regard to any real-world setting. The results sit within our study context, where scaffolding our findings to an ordinary environment is the next logical step and would ground our results. Reflecting further upon this, we find that, although we comment upon gestures and the role they play within the interaction, the role that gestures play within our context may be greater than presented here, and it is arguably beyond the scope of this paper. Future work could investigate these too and develop a gesture set or framework to capture and create continues gestures that support interaction across devices, as represented by the efforts of Dingler et al [7]. As things stand, there is still some distance to go in this field between our research and normal interaction. For not, no clear-cut assumptions can be made.

Looking forward, we suggest that our findings on affordances can ground the design of more novel technologies, thereby improving on starting with *ad hoc* prototypes, which are often developed without a clear understanding of how they might be used. For example, with e-textiles, both worn and Internet of Things devices such as Gaver et al.'s [13] History tablecloth, browsing people and sharing information could necessitate both private and public ways to display information in an open-ended situations. Here, we stress the importance of evaluating current technologies as seeds from which to grow these novel devices.

Additionally, whilst our sample was a mixture of students and professionals, it is unknown whether new relationships, locales, and professions or changes in the scenario context would alter the results significantly. The findings have limited scope for generalisation, as our users' perceptions may differ from those of the general population. On this point, we should stress that one key limitation of our work is the hyperlocal setting of our participants – investigations have yet to be done with regard to how our findings tie in with various societal norms, cultures, and contexts. Also, our study was conducted in a highly technologically advanced country (Finland), and that may have influenced the findings. Both the acceptability of the technological devices and the prior exposure to such devices – e.g., participants having used head-mounted displays and smartwatches – may be relevant in this connection. Still, only 50% and 25%, respectively, had used these, and a novelty factor remains, which could have its own implications, in this context and others, especially with regard to head-mounted displays. It bears remembering also that usage does not imply experience in this case. More investigation is needed to model use and experience in respect of the user's overall experience of relevant devices. Our work also presumed able-bodiedness of the participants, and some facets of the situation would change if participants were to need assistive technology.

As our results indicate, there is value in using the scenario outlined above, which goes beyond existing work, though we still need to expand on the results, generalise them, and understand them in light of other scenarios in several key respects. Proceeding from here, we can point to the profiles curated by the participants curated as being static representations, as used in previous work in this area [19–21, 31, 32]. Therein lies an opportunity to investigate how different types of profiles, such as chat-based, gamified profiles or something creative to support self-expression, affect initiation of social interactions angled toward exploration. Through changing the profiles and possibly adding such interactivity, as with annotation or highlighting of people one has already met, we could explore an interesting aspect of supporting conversations. This would also reduce the cognitive load of remembering details between various entities.

The work described here offers a starting point for further mapping of the face-to-face digital augmentation space toward how such systems can be implemented. This will allow us to not only know what technologies a user would choose but also better consider how the user would appropriate it and the behaviours to support.

## 10 CONCLUDING REMARKS

Our study has provided significant insight into how individuals would employ face-to-face augmentations with smartphones, smartwatches, and head-mounted displays in multiple-user settings, with particular emphasis on how such augmentation would work with a heterogeneous set of interaction devices to view and share the augmentations beyond the existing one-on-one scenarios studied thus far. Our results show that a mix of devices can better support face-to-face interaction, with each device supplying its own means through its particular affordances. These get crystallised nicely once the tensions between the technologies are highlighted. Smartphones support the most affordances, whereas head-mounted displays support the narrowest range of behaviours but were preferred by our participants. Whilst smartwatches were seen as socially seamless, with participants liking the hands-free approach, the small screen and lack of mobility had a heavy impact on their usability. As devices grow so intimately connected to us and our conversations, deciding how we use these devices to augment our interactions, and indeed determining the impact their use will have, is imperative. We hope this project will be aided by the early framework offered here, with its four stages for labelling users' behaviours with digital representations of self for social interactions – *browsing*, *verification*, *viewing*, and *settling* – and by the initial groundwork we have laid for device recommendations both for the interfaces and for the software itself. We have pointed to future directions for continued work to further locate our findings in connection with different relationship, societal, and contextual settings, and we hope these seeds bear fruit.

## ACKNOWLEDGEMENTS

We gratefully thank our study participants, reviewers for their valuable feedback, and Tapio Takala for helping with this project.

## REFERENCES

[1] Aditi Paul. 2019. How are We Really Getting to Know One Another? Effect of Viewing Facebook Profile Information on Initial Conversational Behaviors Between Strangers. The Journal of Social Media in Society, [S.l.], v. 8, n. 1, p. 249-270. ISSN 2325-503x

[2] Deepak Akkil and Poika Isokoski. 2016. Accuracy of interpreting pointing gestures in egocentric view. In Proceedings of the 2016 ACM International Joint Conference on Pervasive and Ubiquitous Computing (UbiComp '16). ACM, New York, NY, USA, 262-273. DOI: https://doi.org/10.1145/2971648.2971687

[3] Susan E. Brennan. 1998. The grounding problem in conversations with and through computers. Social and cognitive approaches to interpersonal communication, pp.201-225.

[4] Marta E. Cecchinato, Anna L. Cox, and Jon Bird. 2015. Smartwatches: the Good, the Bad and the Ugly?. In Proceedings of the 33rd Annual ACM Conference Extended Abstracts on Human Factors in Computing Systems (CHI EA '15). ACM, New York, NY, USA, 2133-2138. DOI: https://doi.org/10.1145/2702613.2732837

[5] Jay Chen and Azza Abouzied. 2016. One LED is Enough: Catalyzing Face-to-face Interactions at Conferences with a Gentle Nudge. In Proceedings of the 19th ACM Conference on Computer-Supported Cooperative Work & Social Computing (CSCW '16). ACM, New York, NY, USA, 172-183. DOI: https://doi.org/10.1145/2818048.2819969

[6] Jay Chen and Azza Abouzied. 2016. One LED is Enough: Catalyzing Face-to-face Interactions at Conferences with a Gentle Nudge. In Proceedings of the 19th ACM Conference on Computer-Supported Cooperative Work & Social Computing (CSCW '16). ACM, New York, NY, USA, 172-183. DOI: https://doi.org/10.1145/2818048.2819969

[7] Tilman Dingler, Rufat Rzayev, Alireza Sahami Shirazi, and Niels Henze. 2018. Designing Consistent Gestures Across Device Types: Eliciting RSVP Controls for Phone, Watch, and Glasses. In Proceedings of the 2018 CHI Conference on Human Factors in Computing Systems (CHI '18). ACM, New York, NY, USA, Paper 419, 12 pages. DOI: https://doi.org/10.1145/3173574.3173993

[8] Brian .L. Due. 2015. The social construction of a Glasshole: Google Glass and multiactivity in social interaction. PsychNology Journal, 13(2).

[9] Barrett Ens, Tovi Grossman, Fraser Anderson, Justin Matejka, and George Fitzmaurice. 2015. Candid Interaction: Revealing Hidden Mobile and Wearable Computing Activities. In Proceedings of the 28th Annual ACM Symposium on User Interface Software & Technology (UIST '15). ACM, New York, NY, USA, 467-476. DOI: https://doi.org/10.1145/2807442.2807449

[10] Ens, B., Quigley, A., Yeo, H.-S., Irani, P., Piumsomboon, T., & Billinghurst, M. (2018). Counterpoint: Exploring Mixed-Scale Gesture Interaction for AR Applications. Extended Abstracts of the 2018 CHI Conference on Human Factors in Computing Systems - CHI '18, 1–6. https://doi.org/10.1145/3170427.3188513

[11] Nicholas Epley and Juliana Schroeder. 2014. Mistakenly seeking solitude. Journal of Experimental Psychology: General 143.5 (2014): 1980.

[12] Shelly D. Farnham and Elizabeth F. Churchill. 2011. Faceted identity, faceted lives: social and technical issues with being yourself online. In Proceedings of the ACM 2011 conference on Computer supported cooperative work (CSCW '11). ACM, New York, NY, USA, 359-368. DOI: https://doi.org/10.1145/1958824.1958880

[13] William Gaver, John Bowers, Andy Boucher, Andy Law, Sarah Pennington, and Nicholas Villar. 2006. The history tablecloth: illuminating domestic activity. In Proceedings of the 6th conference on Designing Interactive systems (DIS '06). ACM, New York, NY, USA, 199-208. DOI: https://doi.org/10.1145/1142405.1142437

[14] Erving Goffman. 1959. The Presentation of Self in Everyday Life. University of Edinburgh, Social Sciences Research Centre. Penguin Books. https://doi.org/10.2307/258197

[15] Sten Govaerts, Adrian Holzer, Bruno Kocher, Andrii Vozniuk, Benot Garbinato, and Denis Gillet. 2018. Blending Digital and Face-to-face Interaction using a Co-located Social Media App in Class. IEEE Transactions on Learning Technologies. DOI: https://doi.org/10.1109/TLT.2018.2856804

[16] Saul Greenberg, Nicolai Marquardt, Till Ballendat, Rob Diaz-Marino, and Miaosen Wang. 2011. Proxemic interactions: the new ubicomp?. Interactions 18, 1 (January 2011), 42-50. DOI: https://doi.org/10.1145/1897239.1897250

[17] Anders Henrysson, Mark Billinghurst, and Mark Ollila. 2005. Face to Face Collaborative AR on Mobile Phones. In Proceedings of the 4th IEEE/ACM International Symposium on Mixed and Augmented Reality (ISMAR '05). IEEE Computer Society, Washington, DC, USA, 80-89. DOI: https://doi.org/10.1109/ISMAR.2005.32

[18] Hindmarsh, Jon, and Christian Heath. 2000. Embodied reference: A study of deixis in workplace interaction. Journal of Pragmatics32, no. 12 (2000): 1855-1878. https://doi.org/10.1016/S0378-2166(99)00122-8

[19] Pradthana Jarusriboonchai, Aris Malapaschas, and Thomas Olsson. 2016. Design and Evaluation of a Multi-Player Mobile Game for Icebreaking Activity. In Proceedings of the 2016 CHI Conference on Human Factors in Computing Systems (CHI '16). ACM, New York, NY, USA, 4366-4377. DOI: https://doi.org/10.1145/2858036.2858298

[20] Pradthana Jarusriboonchai, Aris Malapaschas, Thomas Olsson, and Kaisa Väänänen. 2016. Increasing Collocated People's Awareness of the Mobile User's Activities: a Field Trial of Social Displays. In Proceedings of the 19th ACM Conference on Computer-Supported Cooperative Work & Social Computing (CSCW '16). ACM, New York, NY, USA, 1691-1702. DOI: https://doi.org/10.1145/2818048.2819990

[21] Pradthana Jarusriboonchai, Thomas Olsson and Kaisa Väänänen-Vainio-Mattila. 2014. Opportunities and Challenges of Mobile Applications as "Tickets -to- Talk": A Scenario -Based User Study. In Mum'14 (pp. 89–97).

[22] Pradthana Jarusriboonchai, Thomas Olsson, Vikas Prabhu, and Kaisa Väänänen-Vainio-Mattila. 2015. CueSense: A Wearable Proximity-Aware Display Enhancing Encounters. In Proceedings of the 33rd Annual ACM Conference Extended Abstracts on Human Factors in Computing Systems (CHI EA '15). ACM, New York, NY, USA, 2127-2132. DOI: https://doi.org/10.1145/2702613.2732833

[23] Simon Jones and Eamonn O'Neill. 2011. Contextual dynamics of group-based sharing decisions. In Proceedings of the SIGCHI Conference on Human Factors in Computing Systems (CHI '11). ACM, New York, NY, USA, 1777-1786. DOI: https://doi.org/10.1145/1978942.1979200

[24] Viirj Kan, Katsuya Fujii, Judith Amores, Chang Long Zhu Jin, Pattie Maes, and Hiroshi Ishii. 2015. Social Textiles: Social Affordances and Icebreaking Interactions Through Wearable Social Messaging. In Proceedings of the Ninth International Conference on Tangible, Embedded, and Embodied Interaction (TEI '15). ACM, New York, NY, USA, 619-624. DOI: https://doi.org/10.1145/2677199.2688816

[25] Hsin-Liu (Cindy) Kao and Chris Schmandt. 2015. MugShots: A Mug Display for Front and Back Stage Social Interaction in the Workplace. In Proceedings of the Ninth International Conference on Tangible, Embedded, and Embodied Interaction (TEI '15). ACM, New York, NY, USA, 57-60. DOI: https://doi.org/10.1145/2677199.2680557

[26] John Kennedy & Michael Satran, 2018. Guidelines. Microsoft Dev Center. Microsoft. Avaliable at: https://docs.microsoft.com/en-us/windows/desktop/uxguide/guidelines

[27] Kiyoshi Kiyokawa, Mark Billinghurst, Bruce Campbell, and Eric Woods. 2003. An Occlusion-Capable Optical See-through Head Mount Display for Supporting Co-located Collaboration. In Proceedings of the 2nd IEEE/ACM International Symposium on Mixed and Augmented Reality (ISMAR '03). IEEE Computer Society, Washington, DC, USA, 133-.

[28] Lisa Kleinman, Tad Hirsch, and Matt Yurdana. 2015. Exploring Mobile Devices as Personal Public Displays. In Proceedings of the 17th International Conference on Human-Computer Interaction with Mobile Devices and Services (MobileHCI '15). ACM, New York, NY, USA, 233-243. DOI: https://doi.org/10.1145/2785830.2785833

[29] Marion Koelle, Matthias Kranz, and Andreas Möller. 2015. Don't look at me that way!: Understanding User Attitudes Towards Data Glasses Usage. In Proceedings of the 17th International Conference on Human-Computer Interaction with Mobile Devices and Services (MobileHCI '15). ACM, New York, NY, USA, 362-372. DOI: https://doi.org/10.1145/2785830.2785842

[30] Jaimie Arona Krems, Robin I.M. Dunbar and Steven L. Neuberg. 2016. Something to talk about: are conversation sizes constrained by mental modeling abilities?. Evolution and Human Behavior, 37(6), pp.423-428. DOI: https://doi.org/10.1016/j.evolhumbehav.2016.05.005

[31] Mikko Kytö and David McGookin. 2017. Augmenting Multi-Party Face-to-Face Interactions Amongst Strangers with User Generated Content. Computer Supported Cooperative Work: CSCW: An International Journal, 26(4–6), 527– 562. DOI: http://doi.org/10.1007/s10606-017-9281-1

[32] Mikko Kytö and David McGookin. 2017. Investigating user generated presentations of self in face-to-face interaction between strangers. International Journal of Human ComputerStudies, 104(February), 1–15. DOI: http://doi.org/10.1016/j.ijhcs.2017.02.007

[33] Cliff Lampe, Nicole Ellison, and Charles Steinfield. 2006. A face(book) in the crowd: social Searching vs. social browsing. In Proceedings of the 2006 20th anniversary conference on Computer supported cooperative work (CSCW '06). ACM, New York, NY, USA, 167-170. DOI=http://dx.doi.org/10.1145/1180875.1180901

[34] Andrés Lucero, Jussi Holopainen, and Tero Jokela. 2011. Pass-them-around: collaborative use of mobile phones for photo sharing. In Proceedings of the SIGCHI Conference on Human Factors in Computing Systems (CHI '11). ACM, New York, NY, USA, 1787-1796. DOI: https://doi.org/10.1145/1978942.1979201

[35] Paul Luff and Christian Heath. 1998. Mobility in collaboration. In Proceedings of the 1998 ACM conference on Computer supported cooperative work(CSCW '98). ACM, New York, NY, USA, 305-314. DOI=http://dx.doi.org/10.1145/289444.289505

[36] Nicolai Marquardt, Ken Hinckley, and Saul Greenberg. 2012. Cross-device interaction via micro-mobility and f-formations. In Proceedings of the 25th annual ACM symposium on User interface software and technology (UIST '12). ACM, New York, NY, USA, 13-22. DOI: https://doi.org/10.1145/2380116.2380121

[37] Julia M. Mayer, Starr Roxanne Hiltz, and Quentin Jones. 2015. Making Social Matching Context-Aware: Design Concepts and Open Challenges. In Proceedings of the 33rd Annual ACM Conference on Human Factors in Computing Systems (CHI '15). ACM, New York, NY, USA, 545-554. DOI: https://doi.org/10.1145/2702123.2702343

[38] Julia M. Mayer, Starr Roxanne Hiltz, Louise Barkhuus, Kaisa Väänänen, and Quentin Jones. 2016. Supporting Opportunities for Context-Aware Social Matching: An Experience Sampling Study. In Proceedings of the 2016 CHI Conference on Human Factors in Computing Systems (CHI '16). ACM, New York, NY, USA, 2430-2441. DOI: https://doi.org/10.1145/2858036.2858175

[39] Gerard McAtamney and Caroline Parker. 2006. An examination of the effects of a headmounted display on informal face-to-face communication. In Proceedings of the SIGCHI Conference on Human Factors in Computing Systems (CHI '06).

[40] Rebecca Grinter, Thomas Rodden, Paul Aoki, Ed Cutrell, Robin Jeffries, and Gary Olson (Eds.). An examination of the effects of a wearable display on informal face-to-face communicationACM, New York, NY, USA, 45-54. DOI: http://dx.doi.org/10.1145/1124772.1124780

[41] Joseph F. McCarthy, David W. McDonald, Suzanne Soroczak, David H. Nguyen, and Al M. Rashid. 2004. Augmenting the social space of an academic conference. In Proceedings of the 2004 ACM conference on Computer supported cooperative work (CSCW '04). ACM, New York, NY, USA, 39-48. DOI: http://dx.doi.org/10.1145/1031607.1031615

[42] David W. McDonald, Joseph F. McCarthy, Suzanne Soroczak, David H. Nguyen, and Al M. Rashid. 2008. Proactive displays: Supporting awareness in fluid social environments. ACM Trans. Comput.-Hum. Interact. 14, 4, Article 16 (January 2008), 31 pages. DOI: http://dx.doi.org/10.1145/1314683.1314684

[43] David McGookin. 2014. Digital aura: investigating representations of self in augmented reality applications. In Proceedings of the 8th Nordic Conference on Human-Computer Interaction: Fun, Fast, Foundational (NordiCHI '14). ACM, New York, NY, USA, 1007-1010. DOI: https://doi.org/10.1145/2639189.2670262

[44] Christian Muller-Tomfelde and Morten Fjeld. 2012. Tabletops: Interactive Horizontal Displays for Ubiquitous Computing. Computer 45, 2 (February 2012), 78-81. DOI: https://doi.org/10.1109/MC.2012.64

[45] Wai Shan (Florence) Ng and Ehud Sharlin. 2010. Tweeting halo: clothing that tweets. In Adjunct proceedings of the 23nd annual ACM symposium on User interface software and technology (UIST '10). ACM, New York, NY, USA, 447-448. DOI: https://doi.org/10.1145/1866218.1866264

[46] Tien T. Nguyen, Duyen T. Nguyen, Shamsi T. Iqbal, and Eyal Ofek. 2015. The Known Stranger: Supporting Conversations between Strangers with Personalized Topic Suggestions. In Proceedings of the 33rd Annual ACM Conference on Human Factors in Computing Systems (CHI '15). ACM, New York, NY, USA, 555-564. DOI: https://doi.org/10.1145/2702123.2702411

[47] Michael Nielsen, Moritz Störring, Thomas B. Moeslund, and Erik Granum. 2004. A Procedure for Developing Intuitive and Ergonomic Gesture Interfaces for HCI. Springer Berlin Heidelberg, Berlin, Heidelberg, 409–420. DOI:http://dx.doi.org/10.1007/978-3-540-24598-8_38

[48] Eyal Ofek, Shamsi T. Iqbal and Karin Strauss. 2013. Reducing disruption from subtle information delivery during a conversation: mode and bandwidth investigation. In CHI (pp. 3111–3120). DOI: http://doi.org/10.1145/2470654.2466425

[49] Thomas Olsson, Pradthana Jarusriboonchai and Kaiser Väänänen-Vainio-Mattila. 2015, April. Towards Headmounted displays Aiming to Enhance Social Interaction. In Submission to CHI'15 workshop Mobile Collocated Interactions: From Smartphones to Wearables. Retrieved June (Vol. 11, p. 2015).

[50] Jason Orlosky, Kiyoshi Kiyokawa, and Haruo Takemura. 2013. Dynamic text management for see-through wearable and heads-up display systems. In Proceedings of the 2013 international conference on Intelligent user interfaces (IUI '13). ACM, New York, NY, USA, 363-370. DOI: https://doi.org/10.1145/2449396.2449443

[51] Susanna Paasovaara, Ekaterina Olshannikova, Pradthana Jarusriboonchai, Aris Malapaschas, and Thomas Olsson. 2016. Next2You: a proximity-based social application aiming to encourage interaction between nearby people. In Proceedings of the 15th International Conference on Mobile and Ubiquitous Multimedia (MUM '16). ACM, New York, NY, USA, 81-90. DOI: https://doi.org/10.1145/3012709.3012742

[52] Jennifer Pearson, Simon Robinson, and Matt Jones. 2015. It's About Time: Smartwatches as Public Displays. In Proceedings of the 33rd Annual ACM Conference on Human Factors in Computing Systems (CHI '15). ACM, New York, NY, USA, 1257-1266. DOI: https://doi.org/10.1145/2702123.2702247

[53] Yi-Hao Peng, Ming-Wei Hsi, Paul Taele, Ting-Yu Lin, Po-En Lai, Leon Hsu, Tzu-chuan Chen, Te-Yen Wu, Yu-An Chen, Hsien-Hui Tang, and Mike Y. Chen. 2018. SpeechBubbles: Enhancing Captioning Experiences for Deaf and Hard-of-Hearing People in Group Conversations. In Proceedings of the 2018 CHI Conference on Human Factors in Computing Systems (CHI '18). ACM, New York, NY, USA, Paper 293, 10 pages. DOI: https://doi.org/10.1145/3173574.3173867

[54] Stefania Pizza, Barry Brown, Donald McMillan, and Airi Lampinen. 2016. Smartwatch in vivo. In Proceedings of the 2016 CHI Conference on Human Factors in Computing Systems (CHI '16). ACM, New York, NY, USA, 5456-5469. DOI: https://doi.org/10.1145/2858036.285852

[55] Andrew K. Przybylski and Netta Weinstein. 2013. Can you connect with me now? How the presence of mobile communication technology influences face-to-face conversation quality. Journal of Social and Personal Relationships, 30(3), 237– 246. DOI: http://doi.org/10.1177/0265407512453827

[56] Umar Rashid and Aaron Quigley. 2009. Ambient displays in academic settings: Avoiding their underutilization. International Journal of Ambient Computing and Intelligence (IJACI), 1(2), pp.31-38.

[57] Claudia Roda, and Julie Thomas. 2006. Attention aware systems. Encyclopedia of Human Computer Interaction. IGI Global. 38-44. DOI = https://doi.org/10.1016/j.chb.2005.12.005

[58] Yvonne Rogers. 2006. Moving on from Weiser's Vision of Calm Computing: Engaging UbiComp Experiences. In UbiComp 2006: Ubiquitous Computing (Lecture Notes in Computer Science), Paul Dourish and Adrian Friday (Eds.), Vol. 4206. Springer Berlin Heidelberg, Berlin, Heidelberg, 404–421. https://doi.org/10.1007/11853565

[59] Daniela K. Rosner and Kimiko Ryokai. 2008. Spyn: augmenting knitting to support storytelling and reflection. In Proceedings of the 10th international conference on Ubiquitous computing (UbiComp '08). ACM, New York, NY, USA, 340-349. DOI: https://doi.org/10.1145/1409635.1409682

[60] Virpi Roto and Antti Oulasvirta. 2005. Need for non-visual feedback with long response times in mobile HCI. In Special interest tracks and posters of the 14th international conference on World Wide Web (WWW '05). ACM, New York, NY, USA, 775-781. DOI: https://doi.org/10.1145/1062745.1062747

[61] Franca Rupprecht, Carol Naranjo, Achim Ebert, Joseph Olakumni, and Bernd Hamann. 2019. When Bigger is Simply Better After all: Natural and Multi-Modal Interaction with Large Displays Using a Smartwatch. In Proceedings of the Twelfth International Conference on Advances in Computer-Human Interactions (ACHI 2019). ISBN: 978-1-61208-686-6

[62] Tracoi Ryan and Sophia Xenos. 2011. Who uses Facebook? An investigation into the relationship between the big five, shyness, narcissism, loneliness, and Facebook usage. Comput. Hum. Behav., 27 (5) (2011), pp. 1658-1664. DOI:http://doi.org/10.4018/jaci.2009040104

[63] Harvey Sacks. 1992. Lectures on Conversation. Basil Blackwell, Oxford.

[64] Sarah Sharples, Sue Cobb, Amanda Moody and John R. Wilson. 2008. Virtual reality induced symptoms and effects (VRISE): Comparison of head mounted display (HMD), desktop and projection display systems. Displays, 29(2), pp.58-69.

[65] Marcus Sanchez Svensson and Tomas Sokoler. 2008. Ticket-to-talk-television: designing for the circumstantial nature of everyday social interaction. In: Proceedings of the 5th Nordic Conference on Human-computer Interaction: Building Bridges. NordiCHI'08. ACM, New York, NY, USA, pp. 334–343. DOI: http://doi.org/10.1145/1463160.1463197

[66] Gerry Stahl, Nancy Law, Ulrike Cress, and Sten Ludvigsen. 2014. Analyzing roles of individuals in small-group collaboration processes. International Journal of Computer-Supported Collaborative Learning 9, no. 4 (2014): 365-370.

[67] Tatar, D.G., Foster, G., and Bobrow, D.G. (1991). Design for Conversation: Lessons from Cognoter. In International Journal of Man-Machine Studies, 34, pp. 185-209.

[68] Wayne Weiten, Dana S. Dunn and Elizabeth Yost Hammer. 2014. Psychology applied to modern life: Adjustment in the 21st century. Cengage Learning

[69] Janelle Ward. 2017. What are you doing on Tinder? Impression management on a matchmaking mobile app. Information, Communication and Society 20.11 (2017): 1644-1659. DOI= https://doi.org/10.1080/1369118X.2016.1252412

[70] Jacob O. Wobbrock, Htet Htet Aung, Brandon Rothrock, and Brad A. Myers. 2005. Maximizing the Guessability of Symbolic Input. In CHI '05 Extended Abstracts on Human Factors in Computing Systems (CHI EA '05). ACM, New York, NY, USA, 1869–1872. DOI: http://dx.doi.org/10.1145/1056808.1057043

[71] Katrin Wolf, Anja Naumann, Michael Rohs, and Jörg Müller. 2011. A taxonomy of microinteractions: Defining microgestures based on ergonomic and scenario-Dependent requirements. Human-Computer Interaction (INTERACT '13). Lecture Notes in Computer Science 6946, 559–575.